\documentstyle[amsmath,twoside,fleqn,espcrc2,epsfig]{article}
    
\newcommand{\Zz}{{\ifmmode Z \else $Z$\fi}\ }

\newcommand{\bc}{\begin{center}}
\newcommand{\ec}{\end{center}}

\title{Charmonium Production at Tevatron, HERA and LHC
\thanks{Work partially
supported by CICYT under contract AEN-96/1718 and by an Acci\'on
Especial de la Generalitat Valenciana, code 6223080}}
\author{M. A. Sanchis-Lozano$^{a,b}$\thanks{E-mail: mas@evalvx.ific.uv.es}
\vspace{0.4cm}\\
(a) Departamento de F\'{\i}sica Te\'orica \\
\vspace{0.1cm}
(b) Instituto de F\'{\i}sica Corpuscular (IFIC)\\
Centro Mixto Universitat de Val\`encia-CSIC \\
Dr. Moliner 50, E-46100 Burjassot, Valencia (Spain)}

\begin{document}
\begin{abstract}
Charmonia hadro- and photo-production are analyzed in the framework
of the color-octet model taking into account higher-order effects 
induced by initial-state radiation of gluons. We argue that 
color-octet matrix elements obtained from Tevatron
data might be reconciled with those extracted from
HERA data on inelastic $J/\psi$ photoproduction.
Finally we estimate cross sections for events
with muons from prompt $J/\psi$'s as a background
for the $B$ physics programme at LHC experiments.
 \end{abstract}
\vspace{0.1in}
\maketitle

\section{INTRODUCTION}
Over the last years, charmonium physics has recovered a leading role 
in the challenge of understanding both perturbative and non-perturbative 
aspects of the strong interaction dynamics. The surplus of $J/\psi$
and ${\psi}'$ resonance production found at the Fermilab
Tevatron \cite{fermi1} triggered an intense theoretical activity as
conventional wisdom (i.e. the color-singlet model) failed dramatically
to reproduce experimental data. Such a discrepancy, by more
than an order of magnitude, led 
to the introduction by Braaten and Fleming \cite{fleming} of a new 
production mechanism, the so-called color-octet model (COM), 
later cast into a well defined framework based on an effective
non-relativistic field theory (NRQCD) \cite{bodwin}. However, 
although those non-perturbative parameters introduced in the model
and needed to bring accordance with
experimental data can be rigorously defined in the effective 
theory (as vacuum expectation values of effective field
operators), the possibility of obtaining numerical values
from lattice calculations seems unfortunately still far.
This current theoretical inability means in practice that such
long-distance parameters have to be determined from fits to
experimental data.\par 
On the other hand, there is the founded
conviction that due to the distinct energy scales, 
charmonium production cross section at high transverse momentum
factorizes into a short- and a long-distance
part, the former calculable from perturbative
QCD, the latter parameterized by NRQCD matrix elements expected to be
{\em universal} - i.e. the same for
different production mechanisms and energies. However, numerical
values of some non-perturbative parameters extracted from Tevatron 
\cite{cho2} and HERA \cite{hera}
apparently are in disagreement each other, casting some doubts on
the validity of the COM itself, or at least on the
above-mentioned universality hypothesis.\par
This paper is partially devoted to argue that the claim on the inconsistency
of the NRQCD matrix elements obtained from hadro- and
photo-production may be premature because the {\em effective intrinsic} 
transverse  momentum of partons has been overlooked.\par
Indeed, it was well known long time ago that particle production
in hadron-hadron collisions 
requires the introduction, usually as a phenomenological input,
of a primordial $k_t$ of partons in addition to their
expected Fermi motion \cite{break}. During this Conference, we have
heard Gemme and Lazzeroni
discuss the need of a large intrinsic $k_t$ to reproduce
experimental data on $D$ and $B$ production. In fact, most of the 
{\em effective} transverse
momentum of partons should be attributed to multi-gluon emission by
the initial state. In the literature sometimes this effect is simulated or 
simply parameterized by means of a convolution 
of the hard process cross section with a Gaussian
$k_t$ distribution whose width is an adjustable parameter
\cite{sridhar}. However, strictly speaking the effect is modeled
more reliably by using a Monte Carlo generation of the real 
complexity of the initial-state radiation. 
As in our previous analysis of charmonium hadro-production \cite{mas0} 
we have used PYTHIA \cite{pythia}
managing multi-gluon emission according to an elaborated
algorithm \cite{sjos}.\par

\section{EFFECT OF THE INITIAL RADIATION ON THE $p_t$ DISTRIBUTION}
Let us point out that, technically, the inclusion of
multi-gluon emission by itself hardly changes the overall
cross section for charmonium production obtained from PYTHIA.
However, the smearing caused by the initial-state radiation
is responsible for a {\em migration of events} towards the high-$p_t$
tail of the charmonium distribution, thus enhancing the
differential cross section in the region where experimental points actually
lie. In figure 1 (see Ref. \cite{mas}) this effect is shown, clearly 
pointing out the need for a correction of the results from
Cho and Leibovich \cite{cho2} if a more realistic description of the 
underlying hadronic
complexity of the interaction (i.e. initial-state radiation) is required. 

\begin{figure}[htb]
\begin{center}
\mbox{\epsfig{file=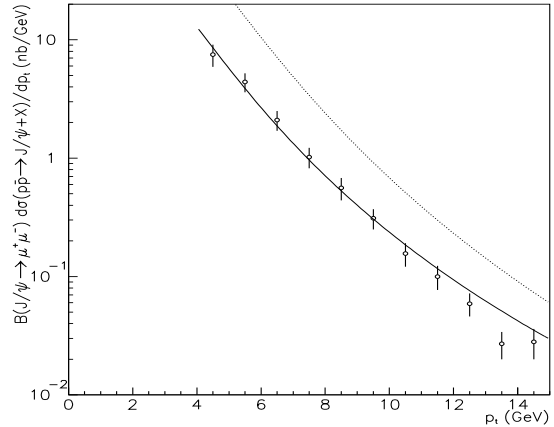,height=6.5cm,width=8.cm}}\end{center}
\caption{Curves obtained from PYTHIA (not fit) including
the color-octet mechanism for $J/\psi$ production at the Tevatron
using the same parameters and parton distribution function as in Ref. [4].
The solid line corresponds
to initial and final state radiation off and the dotted line to initial 
and final state radiation on.} 
\end{figure}

In taking into account this $p_t$-dependent $K$ factor, new (lower) values 
for the long-distance parameters, and ultimately the color-octet
matrix elements, are obtained from charmonium hadroproduction. In Table 1 we 
present our values \cite{mas} determined from older Tevatron data \cite{fermi1}
together with the new ones \cite{fermi2,kniehl} on $J/\psi$ hadroproduction, 
both showing a better agreement \cite{kniehl} with HERA results than 
before \cite{cho2,kniehl}.

\begin{table*}[htb]
\setlength{\tabcolsep}{1.5pc}
\newlength{\digitwidth} \settowidth{\digitwidth}{\rm 0}
\caption{Color-octet matrix elements (in units of GeV$^3$) from 
best fits to old and new 
Tevatron data on prompt $J/\psi$ production. 
($M_r^{J/\psi}=<0{\mid}O_8^{J/\psi}(^1S_0){\mid}0>\ +\ 
r{\cdot}<0{\mid}O_8^{J/\psi}(^3P_0){\mid}0>/m_c^2$.) 
\hrule
}
\label{FACTORES}

\begin{center}
\begin{tabular}{lcc}    \hline
NRQCD matrix element:  & $<0{\mid}O_8^{J/\psi}(^3S_1){\mid}0>$ & 
$M_r^{J/\psi}$ \\
\hline

Ref. \cite{mas} (GRV, $r=3$): &$(3.4{\pm}0.4){\times}10^{-3}$& 
$(6.0{\pm}1.2){\times}10^{-3}$ \\
Ref. \cite{kniehl} ($r=3.5$): & $(2.73{\pm}0.45){\times}10^{-3}$ & 
$(5.72{\pm}1.84){\times}10^{-3}$ \\
\hline
\end{tabular}
\end{center}
\end{table*}

\section{THEORETICAL UNCERTAINTIES}
	There is no space to review all the theoretical uncertainties
related to the extraction of matrix elements from charmonium
production which on the other hand have extensively been reviewed in
the literature (see for example \cite{schuler}). Let us only
mention:
\begin{itemize}
\item Ambiguities in the choice of energy scales, charm mass, PDF, ...
\item Partial breaking of the heavy-quark spin symmetry in charmonia 
systems, ...
\item Subleading $O(>v^7)$ contributions, higher-order ${\alpha}_s$
effects, ...
\item Possible kinematic effects near boundaries of phase space
\item Possible effects due to the surrounding medium breaking universality 
\end{itemize}
\par
Below we comment in some detail on the possible effects due to
the sensitivity to the kinematics of soft
gluon radiation in the long-distance evolution of the
$c\overline{c}$ pair into final charmonium \cite{beneke}.

\subsection{Finite width effects}
	Finite width effects can have consequences
as Ernstr\"{o}m, L\"{o}nnblad and V\"{a}nttinen discussed some time ago 
\cite{vanti}. Indeed,  
even a small energy difference (of order $Mv^2$) between the mass of final 
charmonium and the invariant mass of the created $c\overline{c}$ pair 
should have an effect on the (steep) momentum distribution. Nevertheless, 
the point is that the formation of an intermediate colored $c\overline{c}$ 
state does not preclude the possibility of an energy deficit and not 
necessarily an excess w.r.t. charmonium mass, the 
latter implying the emission of soft gluons as usually considered. In other 
words, the $c\overline{c}$ pair could get energy from the 
surrounding medium. 
\par
This viewpoint is in fact not surprising since in the color 
evaporation model \cite{frit} based on duality arguments, the 
cross section for charmonium production is written as an integral over the
$\hat{s}$ range $4m_c^2$ and $4m_D^2$ (i.e. the upper limit equal to open
charm production). However, charmonium production takes place in a $\lq\lq$hot" 
medium as Hoyer pointed out in his talk (for more
details see Ref. \cite{hoyer}), so sub-thresholds effects likely should
have an influence. In the improved color evaporation model this means
that the heavy quark pair can get energy from the surrounding color field
to give open charm production beneath $\hat{s}=4m_D^2$ \cite{vogt}. 
\par
Analogously in the COM, those dipole transitions occurring 
during the evolution into final particles should be
associated to absorption of soft gluons from the neighborhood as well.
For example a $c\overline{c}[^3S_1^{(8)}]$ pair may absorb two 
non-perturbative gluons
becoming a $O(v^2)$ Fock component of the $J/\psi$ state or may radiate
two soft gluons thus with nearly unity overlap with the $J/\psi$ wave function.
The conclusion is that on the average one should consider both 
possibilities, likely decreasing the net impact of the above-mentioned
effect on the $p_t$ distribution.

\begin{figure}[htb]
\begin{center}
\mbox{\epsfig{file=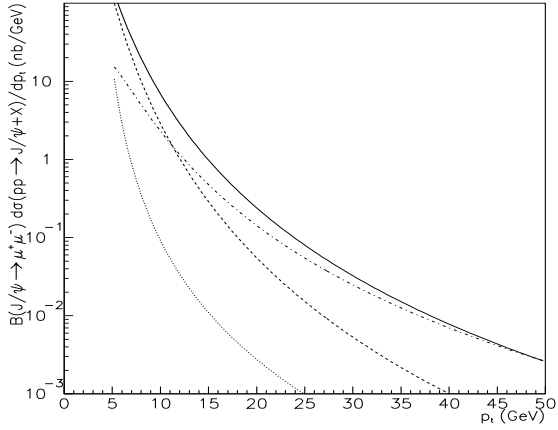,height=6.5cm,width=8.cm}}\end{center}
\caption{Our predictions for prompt $J/\psi$ direct production at the LHC
according to the color-octet model implemented in PYTHIA for the CTEQ
PDF [10]. Dotted line: CSM; dashed line:
$^1S_0^{(8)}+^3P_J^{(8)}$ contributions; dot-dash line: $^3S_1^{(8)}$
contribution; solid line: all contributions. Notice that asymptotically
the $^3S_1^{(8)}$ contribution becomes dominant, likely
implying a transverse polarization of charmonium at high $p_t$.}
\end{figure}

\section{LHC PREDICTIONS}
As outlined in the previous section, the extraction of NRQCD matrix 
elements from charmonium
production is not yet fully satisfactory
due to several theoretical uncertainties. Still more analysis
has to be done to check the universality hypothesis allowing
comparisons between different production reactions. Polarization
of charmonium from gluon fragmentation at high $p_t$ might be decisive
to verify the validity of the COM \cite{beneke2}.\par 
Nevertheless, under the assumption of the validity of the COM
to explain charmonia production, long-distance matrix elements obtained 
from Tevatron data should parameterize those uncertainties to a large
extent. Therefore, in extrapolating to
LHC energies, predictions should be reliable if correctly
performed. In our case, this actually means running PYTHIA with the
same parameters and options (e.g. radiation on) as previously employed
in the fit to Tevatron experimental points.\par 
In Figure 2 we show
our predictions for prompt $J/\psi$ direct production at the LHC
using the CTEQ PDF. A caveat is in order, however. 
In changing the choice for the PDF there are noticeable changes on the 
production rate by about $50\%$ in certain regions of $p_t$ 
(see Ref.\cite{mas}). 

\par

\subsection{Muon cross sections from prompt $J/\psi$ production at
the LHC}
We have estimated the cross sections for events with
muons from prompt $J/\psi$'s, passing typical pseudorapidity
and transverse momentum cuts, foreseen in the first and
second level triggers for $B$ physics at the two LHC experiments
ATLAS and CMS. We shall denote as ${\mu}3$, ${\mu}6$ 
(${\mu}3{\mu}3$, ${\mu}6{\mu}3$) those events with at least one (two)
muon(s) whose $p_t$ is (are) larger than 3 GeV or 6 GeV, respectively. 
\par
From our modified version of PYTHIA with the COM implemented in, we
obtain 
\begin{eqnarray}
\sigma({\mu}3)\ =\  622 & nb & \nonumber \\
\sigma({\mu}6)\ =\ 52 & nb & \nonumber \\
\sigma({\mu}3{\mu3})\ =\  30 & nb & \nonumber \\ 
\sigma({\mu}6{\mu3})\ =\ 9 & nb & \nonumber 
\end{eqnarray}
\par
In all cases an additional pseudorapidity cut ${\mid}{\eta}{\mid}<2.5$
for muons was required on generated events.\par
Those numbers should be useful to evaluate an order-of-magnitude
estimate of background sources
for the $B$ physics programme at LHC experiments, especially regarding
the search for CP violation via the golden decay mode 
$B_d^0{\rightarrow}{J/\psi}K_s$.
\subsubsection*{Acknowledgments}
I thank P. Eerola, N. Ellis, P. Hoyer and M. Smizanska for interesting
discussions.

\end{document}